\documentclass[
 reprint,
 superscriptaddress,
showpacs,preprintnumbers,
 amsmath,amssymb,
 aps,
prl
]{revtex4-1}

\usepackage{amsmath,amssymb,bbm,epsfig}
\usepackage{hyperref}

\usepackage{color}
\usepackage[usenames,dvipsnames]{xcolor}

\newcommand{\be}{\begin{equation}}
\newcommand{\ee}{\end{equation}}
\newcommand{\ba}{\begin{array}}
\newcommand{\ea}{\end{array}}
\newcommand{\bqa}{\begin{eqnarray}}
\newcommand{\eqa}{\end{eqnarray}}

\newcommand{\um}{\mathbbm{1}}

\newcommand{\ket}[1]{\ensuremath{| #1 \rangle}}

\usepackage{pdfpages}

\newcommand{\ie}{{\it i.e. }}
\newcommand{\eg}{{\it e.g. }}

\begin{document}

\title{
Accurate effective Hamiltonians via unitary flow in Floquet space}

\author{Albert \surname{Verdeny}}
\affiliation{Freiburg Institute for Advanced Studies, Albert-Ludwigs-Universit\"at, Albertstrasse 19, 79104 Freiburg, Germany}

\author{Andreas Mielke}
\affiliation{Institut f\"ur Theoretische Physik, Universit\"at Heidelberg, Philosophenweg 19, 69120 Heidelberg, Germany}

\author{Florian Mintert}
\affiliation{Freiburg Institute for Advanced Studies, Albert-Ludwigs-Universit\"at, Albertstrasse 19, 79104 Freiburg, Germany}

\begin{abstract}
We present a systematic construction of effective Hamiltonians of periodically driven quantum systems.
Because of an equivalence between the time dependence of a Hamiltonian and an interaction in its Floquet operator,
flow equations, that permit to decouple interacting quantum systems, allow us to identify time-independent Hamiltonians for driven systems.
With this approach, we explain the experimentally observed deviation of expected suppression of tunneling in ultracold atoms.
\end{abstract}

\date{\today}

\maketitle

The idea to use well-controllable quantum systems for simulations to explore physical phenomena has created big expectations to answer questions that exceed our computational and analytical means.
In particular, strongly correlated many-body states, as they typically occur in flat-band systems \cite{Mielke91,Yao12,Moller12}, readily push us to our limitations of classical simulations. 

Suitably driven systems can be used to mimic the dynamics of essentially any Hamiltonian,
but the precise identification of such an effective Hamiltonian for given driving parameters is a big theoretical challenge.
This can be exemplified by the dramatic increase of complexity in going from the static to the driven two-level system.
The former is a standard textbook toy model; the latter is exactly solvable only in a few exceptional cases \cite{Barnes12}.
The effective Hamiltonian of a given driven system is typically found in an approximate manner. The deviations between the actual and approximated dynamics accumulate in time and become significant for sufficiently long times. In order to perform precise quantum simulations it is therefore crucial to develop tools that allow one to systematically construct effective Hamiltonians with high accuracy.

As a prominent example of driving-induced effective dynamics, we highlight shaken optical lattices \cite{Eckardt05,Lignier07}, which permit the engineering of the tunneling of ultracold bosonic atoms confined in an optical lattice by appropriately adjusting the driving parameters. This yields \eg dynamical localization \cite{Lignier07,Eckardt09} and provides a promising route towards the simulation of artificial gauge fields \cite{Struck12,Hauke12}.
However, despite the proven success of the usually considered effective Hamiltonian for high driving frequencies, substantial deviations from theoretically predicted dynamical localization have been observed in many-body samples for moderate driving frequencies \cite{Lignier07}, which is in striking contrast to the single-particle case, where the exact dynamical localization occurs irrespectively of the driving frequency \cite{Kayanuma08,Dunlap86}.

In this Letter, we introduce a new approach to derive effective Hamiltonians merging the concepts of Floquet theory \cite{Floquet83,Schirley65} and flow equations (FE) \cite{Wegner94}. With this, we provide an explanation for the  experimentally observed deviations \cite{Lignier07}  from the theoretically predicted  suppression of tunneling \cite{Eckardt05}.

The starting point to arrive at an effective Hamiltonian is a periodically driven Hamiltonian $H(t)=H(t+T)$.
After full cycles of the driving, {\it i.e.} $t=nT$ with integer $n$,
the time-evolution operator, $U(t)={\cal T}\exp\big( -i\int_0^t H(t')dt'\big)$ (${\cal T}$ denotes the time-ordering operator), can be written as $U(nT)=e^{-iH_{\rm eff}nT}$, which defines the effective Hamiltonian $H_{\rm eff}$ \footnote{The effective Hamiltonian is uniquely defined up to multiples of the driving frequency $\omega=2\pi/T$ in its eigenvalues and some other subtleties discussed in Ref. \cite{Gesztesy81}.}.
The actual dynamics will follow the dynamics $U_{\rm eff}(t)=e^{-iH_{\rm eff}t}$ induced by the effective Hamiltonian only stroboscopically,
but in the regime of fast driving, where $\omega$
exceeds the relevant scales of  $H(t)$, the effective dynamics is a good approximation also for $t\neq nT$.
The deviation between exact and effective dynamics defines the unitary
\begin{equation}\label{separationscales}
U_F(t)=U_{\rm eff}(t)U^\dagger (t)\ .
\end{equation}
Since $U(t)$ coincides with $U_{\rm eff}$ at $t=nT$, $U_F(t)$ is periodic with period $T$, and equals the identity at multiples of the period. 

Given Eq. (\ref{separationscales}), the effective Hamiltonian can be obtained from the Schr\"odinger equation, $i(\partial_t U(t))=H(t)U(t)$ ($\hbar=1$), and reads
\begin{equation}
H_{\rm eff}=U_F(t)H(t)U_F(t)^\dagger-iU_F(t)(\partial_tU_F(t)^\dagger).
\end{equation}
Thus, the effective Hamiltonian is found after performing a periodic time-dependent unitary transformation, $U_F(t)$, such that the resulting transformed Hamiltonian is time independent and $U_F(0)=\um$.
In practice, finding this exact unitary transformation is an extremely difficult task.
Here we present a method, using an unconventional approach, that allows one to systematically obtain the effective Hamiltonian up to a required accuracy.

We use the framework of Floquet theory  which asserts that the Schr\"odinger equation with a time-periodic Hamiltonian $H(t)=H(t+T)$
has a complete set of solutions $\ket{\phi_k(t)}$ that decompose into a phase factor and a time-periodic state vector, {\it i.e.}
$\ket{\phi_k(t)}=e^{i\epsilon_k t}\ket{u_k(t)}$ with $\ket{u_k(t)}=\ket{u_k(t+T)}$.
Because of their periodicity, the state vectors $\ket{u_k(t)}$ can be expanded in a discrete set of periodic functions $f_n(t)$,
which are vectors in the space of time-dependent functions defined in the interval $[0,T)$.
In the following we will use the functions $e^{in\omega t}$ as basis and associate with each such function a state vector $\ket{n}$ in a Hilbert space ${\cal H}_T$.

Any T-periodic operator $A(t)=\sum_nA_ne^{in\omega t}$ can now be mapped to an operator in `Floquet space'
\be \label{map}
{\cal A}=\sum_{n}A_n\otimes\sigma_n,
\ee
where the Fourier components $A_n=\frac{1}{T}\int_0^T A(t)e^{-in\omega t}$ act on the Hilbert space of the actual system, and the $\sigma_n$ acting on ${\cal H}_T$ are defined by $\sigma_m\ket{n}=\ket{n+m}$\footnote{
Applying the `operator' $e^{in\omega t}$ on a `wave function' $e^{im\omega t}$, yields
$e^{i(n+m)\omega t}$, {\it i.e.} raises the quantum number $m$ by $n$.}. 
Similarly, the time derivative $-i\partial_t$ is associated with
\be
{\cal D}= \um\otimes \omega \hat{n},
\ee
with the number operator $\hat{n}\ket{n}=n\ket{n}$,
so that the Floquet operator $K(t)=H(t)-i\partial_t$ is mapped to 
\bqa
{\cal K}&=&\sum_{n}H_n\otimes\sigma_n+\um\otimes \omega\hat{n}\\
&=&\underbrace{H_0\otimes\um+\um\otimes \omega\hat{n}}_{{\cal K}_0}+\underbrace{\sum_{n\neq 0}H_n\otimes\sigma_n}_{{\cal K}_{\rm int}} . \label{interactingK}
\eqa
Formally, this is equivalent to a time-independent Hamiltonian of a composite system with a Hamiltonian ${\cal K}_0$ for the individual components and an interaction ${\cal K}_{\rm int}$.
Because of this analogy, techniques to treat interactions in time-independent Hamiltonians are applicable.

If $H(t)$ was time independent all Fourier components $H_n$ but the static $H_0$ would vanish,
so that the interaction term ${\cal K}_{\rm int}$ would vanish as well.
Consequently, a static Hamiltonian is equivalent to a noninteracting system in the present framework.
Our goal is, therefore, to find an operator ${\cal U}_c$ that corresponds to a periodic unitary transformation $U_c(t)$ according to Eq.~\eqref{map},
such that the transformed Floquet operator describes two noninteracting systems, ${\cal U}_c{\cal K}{\cal U}_c^\dagger=H_c\otimes \um+\um\otimes \omega \hat{n}$.
Once such a transformation is found, the sought transformation reads $U_F(t)=U_c^\dagger (0)U_c(t)$,
and the effective Hamiltonian is given by
$H_{\rm eff}=U_c^\dagger(0)H_c U_c(0)$. 

We target the required block diagonalization of the Floquet operator with the method of flow equations \cite{Wegner94,Glazek93},
which is considered a generalization of conventional scaling approaches and is based on a unitary flow that makes the Hamiltonian increasingly diagonal \cite{Wegner06,Kehrein}. 
The method defines a family of unitarily equivalent time-independent Hamiltonians related to each other by a continuous parameter $l$
\begin{equation}\label{flowequation}
\dfrac{dH(l)}{dl}=[\eta(l),H(l)],
\end{equation}
where $\eta(l)$ is the anti-Hermitian generator, $\eta(l)^\dagger=-\eta(l)$, of a unitary transformation.
The boundary conditions are such that $H(l=0)$ coincides with the given Hamiltonian,
and $\eta(l)$ needs to be chosen such that $H(l\rightarrow \infty)$ is in the desired form, {\it i.e.} typically diagonal or block diagonal.
The canonical approach \cite{Wegner94} to eliminate an interaction $H_{\rm int}$ of a Hamiltonian $H=H_0+H_{\rm int}$ is to define
the flowing Hamiltonian $H(l)=H_0(l)+H_{\rm int}(l)$ and the corresponding generator $\eta(l)=[H_0(l), H_{\rm int}(l)]$. The main advantage of the FE method is that it permits an equal treatment of different energy scales in a renormalization formalism, and
a focus on a special regime, e.g. low lying excitations, is not necessary. This is important if one wants to study dynamical properties in nonequilibrium situations \cite{Kehrein05,Moeckel08}.

The FE method is typically used to decouple an interacting many-body system, \textit{e.g.} the spin from the bosonic bath in the spin-boson model \cite{Kehrein98}.
Here, on the other hand,  we will use it to remove the interaction ${\cal K}_{\rm int}$ in Eq. (\ref{interactingK}).
For this purpose, we will define a  flowing Floquet operator ${\cal K}(l)={\cal K}_0(l)+{\cal K}_{\rm int}(l)$
and apply Eq.~\eqref{flowequation} analogously.
Additional care is however necessary in the choice of generator to ensure that the unitary transformation ${\cal U}_c$ corresponds indeed to a periodic time-dependent transformation.
This is the case exactly if ${\cal U}_c$ is invariant under the symmetry transformation ${\cal S}=\um \otimes \sigma_1$, \ie ${\cal S}{\cal U}_c{\cal S}^\dagger={\cal U}_c$, and exactly generators of the form $\sum_n \eta_n(l)\otimes\sigma_n$ preserve this property.

The generator of interest for our purposes reads $[{\cal D},{\cal K}_{\rm int}(l)]$. 
This generator will induce a flow (dynamics with the flowing parameter $l$) that comes to an end if the interaction commutes with $\mathcal{D}$. This, in turn, implies that the interaction is trivial in $\mathcal{H}_T$, \ie it is of the form $\tilde{H}_0\otimes \um$ and the decoupling has been achieved.
Eq.~\eqref{flowequation} defines an infinite set of non-linear differential equations, so that an exact solution can be found only in very exceptional cases.
We will therefore strive for a high-frequency expansion, where this set of equations is truncated at a given power in $1/\omega$.
This requires a modification of the generator \cite{Kehrein} as discussed in section I in the supplementary material. 
In section II we also discuss the driven two-level system for explanatory purposes and reproduce \cite{Rahav03} the energy shift up to fourth order in $1/\omega$.
Here, however, we will focus on the shaken optical lattice in order to address the above mentioned question of suppressed tunneling.

The Hamiltonian of the one-dimensional shaken optical lattice can be written, in the co-moving reference frame \cite{Eckardt05}, as  $H(t)=H_s+H_d(t)$, with the Bose-Hubbard model $H_s=\sum_i J(c_i^\dagger c_{i+1}+c_{i+1}^\dagger c_{i})+U\sum_i \hat{n}_i(\hat{n}_i-1)$ (with periodic boundary conditions) and an additional driving term $H_d(t)=K\cos(\omega t)\sum_i i \hat{n}_i$ that describes the shaking. 
$J$ denotes the hopping matrix element between nearest-neighbor sites and $U$ is the on-site interaction energy. The operators $c_i^{(\dagger)}$ are the usual bosonic annihilation (creation) operators satisfying $[c_i,c_j^\dagger]=\delta_{ij}$ and $\hat{n}_i=c_i^\dagger c_i$. 
In lowest order in $1/\omega$ we find the desired unitary $U_F(t)$ to read
$U_{F}^{(0)}(t)=\exp\big(i\frac{K}{\omega}\sin(\omega t)\sum_j j \hat{n}_j\big)$, so that the transformed
 Hamiltonian reads
\begin{eqnarray}\label{transformedH}
\tilde{H}(t)&=&H_{\rm eff}^{(0)}+\delta H(t),
\end{eqnarray}
with the previously known \cite{Eckardt05} effective Hamiltonian $H_{\rm eff}^{(0)}=\sum_i J^{\rm eff}(c_i^\dagger c_{i+1}+c_{i+1}^\dagger c_{i})+U\sum_i \hat{n}_i(\hat{n}_i-1)$ and $\delta H(t)=\sum_i (\delta^+(t)c_i^\dagger c_{i+1}+\delta^-(t)c_{i+1}^\dagger c_{i})$.
The effective hopping matrix element reads $J^{\rm eff}=J\mathcal{J}_0(K/\omega)$ in terms of the zeroth order Bessel function and $\delta^\pm(t)=J(e^{\pm i \frac{K}{\omega}\sin(\omega t)}-\mathcal{J}_0(K/\omega))$  is a small deviation.
This effective Hamiltonian, $H_{\rm eff}^{(0)}$, is a good approximation for sufficiently large driving, $\omega\gg\{J,U\}$ \cite{Eckardt05}.
Experimentally, however, deviations from this limiting case have been observed and, as we shall see, a systematic improvement of $H_{\rm eff}$ for a finite driving frequency $\omega$ permits to explain these deviations very well.

With Eq.~\eqref{transformedH} as starting point, our approach yields
the effective Hamiltonian
$H_{\rm eff}=H_{\rm eff}^{(0)}+H_{\rm eff}^{(1)}+O(1/\omega^2)$ with
\begin{eqnarray}\label{heff1}
H_{\rm eff}^{(1)}&=&2\beta(K/\omega)\dfrac{U J}{\omega}\sum_{i} c_i^\dagger (\hat{n}_i-\hat{n}_{i+1}) c_{i+1}+H.c. \ ,
\end{eqnarray}
including effects $\sim 1/\omega$, as described in more detail in section III of the supplementary material. 
As a qualitative change as compared to the lowest order effective Hamiltonian $H_{\rm eff}^{(0)}$,
there is tunneling interaction dependent on site occupation,
whose rate $\beta(K/\omega)=2\sum_{m=1}^{\infty}\mathcal{J}_{2m-1}(K/\omega)/(2m-1)$ is given in terms of the $m$th order Bessel functions, $\mathcal{J}_m$. 
In contrast to the tunneling term or the on-site interaction term of the Bose-Hubbard Hamiltonian,
$H_{\rm eff}^{(1)}$ cannot be diagonalized through a suitable choice of single-particle basis, which makes it difficult to develop a simple physical interpretation.
The effect of $H_{\rm eff}^{(1)}$ is probably best demonstrated by its action on a Fock state $|\textbf{n} \rangle$ with well-defined particle number on each lattice site.
For such a state, the operators $\hat{n}_i$
reduce to scalars ($\hat{n}_i|\textbf{n} \rangle=n_i|\textbf{n} \rangle$), such that
$H_{\rm eff}^{(1)}$ reduces to a regular tunneling term with site-dependent tunnel rates.
For a Fock state $\ket{\textbf{n}_0}$ with site-independent particle numbers, this rate is even site independent
$\sum_{i} (c_i^\dagger (\hat{n}_i-\hat{n}_{i+1}) c_{i+1}+h.c.)|\textbf{n}_0\rangle = \sum_i(c_i^\dagger c_{i+1}-c_{i+1}^\dagger c_{i})|\textbf{n}_0\rangle$,
but only depends on the direction of tunneling.
That is, $H_{\rm eff}^{(1)}$ enhances the tunneling in one direction and suppresses it in the other direction.
The directionality is determined by the sign of the driving amplitude
$K$ and the interaction energy $U$ as well as by the specific value of $K/\omega$. On the other hand, for a Fock state $|\textbf{n}_d\rangle$ with a large particle difference between adjacent sites the tunneling rate depends on the particle gradient, $\sum_{i} (c_i^\dagger (\hat{n}_i-\hat{n}_{i+1}) c_{i+1}+h.c.)|\textbf{n}_d\rangle \approx \sum_i(c_i^\dagger c_{i+1}+c_{i+1}^\dagger c_{i}) |\textbf{n}_d\rangle(n_i-n_{i+1})$. In fact, similarly as above the driving parameters can be tuned such that the rate for tunneling events towards highly populated sites is enhanced, which is impossible for the usual kinetic term of the Bose-Hubbard model.

An advantage of the flow equation method over other methods like perturbation theory is its systematics. In section IV and V of the online supplementary material we explicitly show how this method permits, with little extra effort, the identification of the effective Hamiltonian in first order in one of the parameters but in all orders in the other one. In the large interaction energy regime, $J\ll \{U,\omega\}$, we obtain the effective Hamiltonian
\begin{eqnarray}\label{Heffu}
H_{\rm eff}^U&=&H_{\rm eff}^{(0)}-\sum_{n=1}^\infty J\dfrac{U^n}{\omega^n} (\beta^+_n\hat{C}_n^++\beta^-_n\hat{C}_n^-)\nonumber\\
&+&O(J^2/\omega)
\end{eqnarray}
and in the large tunneling regime, $U\ll \{J,\omega\}$,
\begin{eqnarray}\label{Heffj}
H_{\rm eff}^J&=&H_{\rm eff}^{(0)}-U\sum_{n=1}^\infty \mathcal{J}_0(K/\omega)^{n-1}\dfrac{J^n}{\omega^n} (\beta^+_n\hat{T}_n^++\beta^-_n\hat{T}_n^-)\nonumber\\
&+&O(U^2/\omega),
\end{eqnarray}
 where 
\begin{equation} \label{betan}
 \beta_n^\pm(K/\omega)=(\pm 1)^n\sum_{m=1}^\infty \mathcal{J}_m(K/\omega)\dfrac{1+(-1)^{m-n}}{m^n}
\end{equation} 
and  the operators $\hat{C}_n^\pm$ and $\hat{T}_n^\pm$ are  defined recursively via the relations
 \begin{eqnarray}
\hat{C}_1^\pm =\hat{T}_1^\pm &=&[\sum_i c_i^\dagger c_{i\pm 1},\sum_j \hat{n}_j(\hat{n}_j-1)],\\
\hat{C}_{n+1}^{\pm}&=&\big[ \hat{C}_n^\pm,\sum_j \hat{n}_j(\hat{n}_j-1)\big],\\
\hat{T}_{n+1}^{\pm}&=&\big[ \hat{T}_n^\pm,\sum_j (c_j^\dagger c_{j+1}+c_{j+1}^\dagger c_j)\big].
 \end{eqnarray}
Eq. (\ref{heff1}) is obtained from the first term of the series in Eq. (\ref{Heffu}) and (\ref{Heffj}) using $\beta_1^\pm=\pm\beta$ and $\hat{C}_1^\pm=\hat{T}_1^\pm=-2\sum_i c_i^\dagger(\hat{n}_i-\hat{n}_{i\pm 1})c_{i\pm 1}$. 

For non-interacting particles, $H_{\rm eff}^{(1)}$ and all higher order terms vanish and our result confirms \cite{Kayanuma08,Dunlap86} that  $H_{\rm eff}=\sum_i J^{\rm eff}(c_i^\dagger c_{i+1}+c_{i+1}^\dagger c_{i})$ is the exact effective Hamiltonian independently of the value of $J/\omega$. Thus, for noninteracting particles,  the exact suppression of tunneling is expected at multiples of the driving period whenever $K/\omega$ coincides with the zeros of $\mathcal{J}_0$ even for slow driving.

With interacting particles, however, the exact suppression of the tunneling is not possible in general, since $H_{\rm eff}^{(1)}$ and all  higher order terms do not necessarily vanish, and only an approximate suppression in the large-frequency regime can be obtained.
For moderate driving frequencies $\omega\lesssim 2J$ \cite{Lignier07}, $H_{\rm eff}^{(1)}$ is of comparable magnitude as $H_{\rm eff}^{(0)}$ 
and, if $J_{\rm eff} \simeq 0$, $H_{\rm eff}^{(1)}$ describes the dominant tunneling mechanism that can no longer be considered a small higher order correction.
In particular, as shown in Fig. 1, the tunneling rate proportional to $\beta$ is close to maximal for driving amplitudes at which the rate $J\mathcal{J}_0$ vanishes.
This explains the recently experimentally observed deviations from predictions based on $H_{\rm eff}^{(0)}$ that are particularly pronounced when the effective tunneling is expected to be vanishing.
For slower drivings, also higher order terms get more and more important.
Eq. (\ref{betan}) predicts that not only is $\beta$ maximal when $J\mathcal{J}_0$ vanishes, but all $|\beta^\pm_{2n-1}|$ show the same property.
Also the even coefficients are finite (though nonmaximal) when $J\mathcal{J}_0$ vanishes, but rapidly tend to $0$ for increasing $n$.

\begin{figure}[h]
\includegraphics[scale=0.65]{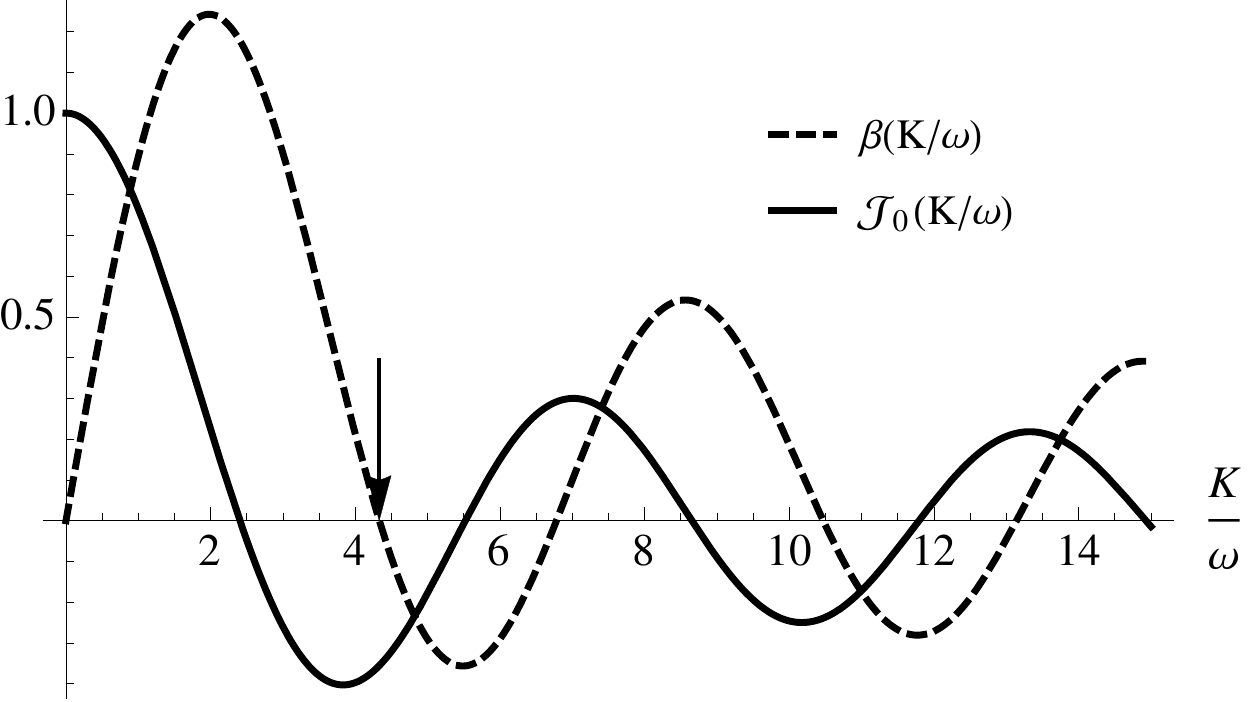}
\caption{\label{plot}
Rates for nearest-neighbor tunneling and tunneling interaction as function of the driving parameter $K/\omega$.
Nearest-neighbor tunneling is proportional to the Bessel function ${\cal J}_0(K/\omega)$ (solid line).
The higher order process of  tunneling interaction [see Eq.~\eqref{heff1}] has a rate proportional to $\beta(K/\omega)$ (dashed line).
For values $K/\omega$ where ${\cal J}_0(K/\omega)=0$, so that the tunneling is expected to be suppressed, the rate for the tunneling interaction is close to maximal, and vice versa.
Particularly clean realizations of Bose-Hubbard models are obtained for $\beta=0$,
 as indicated with an arrow for $K/\omega\approx 4.3$.}
\end{figure}

An enhancement of the tunneling rate would also appear in a more accurate description of the optical lattice system than the Bose-Hubbard model, where next-nearest-neighbor tunneling is neglected assuming that the trapping potential is sufficiently deep \cite{Jaksch98,Eckardt09}.
$H_{\rm eff}^{(1)}$, however, appears also for deep lattices and, as argued above, its signatures can be experimentally observed even with a deep trapping potential, where the Bose-Hubbard model is an excellent approximation. This shows that particular care is required in the identification of effective Hamiltonians and that supposedly high-order contributions can take over a dominant role.

The importance of accurate effective Hamiltonians and their derivation with the present method are by no means limited to the suppression of tunneling discussed above.
For example, in flat band systems a small interaction term results in the emergence of strongly correlated states \cite{Mielke91a},
and only the correct identification of such a seemingly small correction to the effective Hamiltonian will permit the correct prediction of such states.

Since the FE have proven very valuable in the treatment of nonperturbative effects, they bear great potential for situations in which a clear separation of scales is no longer valid. That is, the method presented here can deal with driven strongly interacting systems and identify driving schemes that simulate \eg three-particle interactions. 

As a final remark, we would like to mention that our approach shows an equivalence between the time-independent and the time-dependent FE \cite{Tomaras11}. In our treatment, we use the framework of time-independent FE to treat time-dependent systems. Translating our analysis from Floquet space back to the framework of time-dependent operators, one neatly reproduces the formalism of time-dependent FE. 
This equivalence is completely general and not restricted to periodic Hamiltonians, since one can always treat the time window of interest as the fundamental period of driving.
That is, our approach also permits to translate all the existing expertise on generators of time-independent FE to the much less mature field of time-dependent FE.
In particular, the time-dependent canonical generator \cite{Tomaras11} appears as a natural extension of the time-independent canonical one.

Financial support by the European Research Council within the project ODYCQUENT is gratefully  acknowledged.

\bibliography{mybib}

\begin{widetext}
\clearpage


\section*{Supplementary material}

# Supplementary material

Here, we will sketch how flow equations can be used to find transformations $U_F(t)$ that yield effective Hamiltonians. The starting point is the Floquet operator defined in Eq. (6) in the main paper

$$\mathcal{K} = H_0 \otimes \mathbb{1} + \mathbb{1} \otimes \omega \hat{n} + \sum_{n \neq 0} H_n \otimes \sigma_n$$
$$= \mathcal{H}_0 + \mathcal{D} + \mathcal{K}_{\text{int}}, \tag{1}$$

where $\mathcal{H}_0 = H_0 \otimes \mathbb{1}$, $\mathcal{D} = \mathbb{1} \otimes \omega \hat{n}$ and $\mathcal{K}_{\text{int}} = \sum_{n \neq 0} H_n \otimes \sigma_n$. The largest energy scale of the system is $\omega$, *i.e.* the prefactor of $\mathcal{D}$, and we seek a transformation such that the interactions in the transformed Floquet operator are of order $1/\omega^n$ with a predefined power $n$. In order to do so, we need a properly parametrized flowing Floquet operator $\mathcal{K}(l)$ and generator $\eta(l)$, which can be obtained in an iterative manner as described in the following.

We start with an expansion of the flowing operators

$$\mathcal{H}_0(l) = \sum_{i=1}^{n} a_i(l) \mathcal{O}_i^{(0)} \tag{2}$$

$$\mathcal{K}_{\text{int}}(l) = \sum_{i=1}^{m} b_i(l) \mathcal{O}_i^{(\text{int})} . \tag{3}$$

in a suitably chosen set of single system operators $\{\mathcal{O}_i^{(0)}\}$ and interaction operators $\{\mathcal{O}_i^{(\text{int})}\}$. In order to keep the problem tractable, this set is chosen rather small. In particular, it is *incomplete*. The first ansatz for the generator reads

$$\eta(l) = [\mathcal{D}, \mathcal{K}_{\text{int}}(l)] . \tag{4}$$

The flow equation defines differential equations for the parameters $a_i(l)$ and $b_i(l)$ and can be solved in lowest order in $1/\omega$, that is under the omission of subordinate terms, provided that the parametrization is chosen well. Essentially always, however, is a truncation necessary due to the expansion into an incomplete set of operators $\{\mathcal{O}_i^{(0)}\}$ and $\{\mathcal{O}_i^{(\text{int})}\}$. An inspection of the terms that are dropped permits to improve the parametrization systematically: any contribution to the flow equation that can not be spanned by $\{\mathcal{O}_i^{(0)}\}$ and $\{\mathcal{O}_i^{(\text{int})}\}$ defines a new operator that can be added to these sets. The parametrization of $\mathcal{H}_0(l)$ in Eq. (2) then takes a more general form (the sum includes more terms).

A more general ansatz for the generator

$$\eta'(l) = [\mathcal{D}, \mathcal{K}_{\text{int}}(l) + \sum_i c_i(\{a_i(l), b_i(l)\}) \mathcal{O}_{i+m}^{(\text{int})}] \tag{5}$$

with suitably chosen functions $c_i$ permits to solve the flow equation with a truncation in a higher order in the $1/\omega$ expansion than possible before. This process of extending the sets $\{\mathcal{O}_i^{(0)}\}$ and $\{\mathcal{O}_i^{(\text{int})}\}$ can finally be repeated until the desired accuracy is reached. In fact, interaction-like terms of $n$-th order can be typically neglected if an effective Hamiltonian in $n$-th order is sought, because a further refinement of the generator that allows to solve the flow equations in $n$-th order would result in an $n+1$-st order correction to $\mathcal{H}_0(l)$ only [1]. We will explicitely verify this in the specific cases discussed below.

In order to describe this procedure in more detail, we apply this formalism to the driven two-level system in section I and show that it neatly reproduces the known high-frequency expansion [2]. In section II, we demonstrate that this method indeed permits to identify effective Hamiltonians with high accuracy also for interacting many-body systems.

## I. DRIVEN TWO-LEVEL SYSTEM

The driven two-level system is described by the Hamiltonian

$$H(t) = \omega_0 P_z + 2\Omega \cos(\omega t) P_x, \tag{6}$$

where $P_i$, with $i = x, y, z$, satisfy the commutation relation $[P_x, P_y] = iP_z$ and cyclic permutations. The Floquet operator, $\mathcal{K}$, associated with $H(t)$ reads

$$\mathcal{K} = \omega_0 P_z \otimes \mathbb{1} + \mathbb{1} \otimes \omega \hat{n} + \Omega P_x \otimes (\sigma_1 + \sigma_{-1}). \tag{7}$$

The operators $\sigma_n$ on the second component, defined in Eq. (3) in the main paper, satisfy the relations $\sigma_n \sigma_m = \sigma_{n+m}$, $[\sigma_n, \sigma_m] = 0$ and $[\hat{n}, \sigma_n] = n\sigma_n$. We assume the regime $\omega \gg \{\omega_0, \Omega\}$ and choose the ansatz

$$\mathcal{K}(l) = a(l)P_z \otimes \mathbb{1} + \mathbb{1} \otimes \omega \hat{n} + b(l)P_x \otimes (\sigma_1 + \sigma_{-1}) \tag{8}$$

with the boundary conditions $a(0) = \omega_0$ and $b(0) = \Omega$. There are only two flowing variables $a(l)$ and $b(l)$, that is, the sets $\{\mathcal{O}_i^{(0)}\}$ and $\{\mathcal{O}_i^{(\text{int})}\}$ are comprised only of the single elements $P_z \otimes \mathbb{1}$ and $P_x \otimes (\sigma_1 + \sigma_{-1})$ respectively.

With the generator

$$\eta_1(l) = [\mathcal{D}, \mathcal{K}_{\text{int}}(l)] = \omega b(l) P_x \otimes (\sigma_1 - \sigma_{-1}), \tag{9}$$

the flow equation reads

$$\frac{1}{\omega^2} \frac{d\mathcal{K}(l)}{dl} = -b(l)P_x \otimes (\sigma_1 + \sigma_{-1}) - \frac{i}{\omega} a(l)b(l) P_y \otimes (\sigma_1 - \sigma_{-1}). \tag{10}$$

Eq. (10) can only be solved if the first order term $\frac{i}{\omega}a(l)b(l)P_y \otimes (\sigma_1 - \sigma_{-1})$ is neglected, since it can not be expanded in term of the operators considered hitherto. In this lowest order, one obtains the differential equation $\frac{db(l)}{dl} = -\omega^2 b(l)$, which yields an exponential decay of $b$ with $l$, such that the interaction vanishes for $l \to \infty$ as expected. Nevertheless, as argued above, one will obtain a better approximation to the exact flow equations, if the term $P_y \otimes (\sigma_1 - \sigma_{-1})$ is added to the set $\{\mathcal{O}_i^{(\text{int})}\}$.

With the modified generator

$$\eta_2(l) = \omega b(l) P_x \otimes (\sigma_1 - \sigma_{-1}) + c_1(l) P_y \otimes (\sigma_1 + \sigma_{-1}), \tag{11}$$

the flow equation becomes

$$\frac{1}{\omega^2} \frac{d\mathcal{K}(l)}{dl} = -b(l)P_x \otimes (\sigma_1 + \sigma_{-1}) - \frac{1}{\omega}(ia(l)b(l) + c_1(l))P_y \otimes (\sigma_1 - \sigma_{-1})$$
$$- 2\frac{i}{\omega^2} c_1(l)b(l)P_z \otimes \mathbb{1} + \frac{i}{\omega^2} c_1(l)a(l)P_x \otimes (\sigma_1 + \sigma_{-1}) - \frac{i}{\omega^2} c_1(l)b(l)P_z \otimes (\sigma_2 + \sigma_{-2}). \tag{12}$$

With the choice $c_1(l) = -ia(l)b(l)$ all the first order ($\sim \frac{1}{\omega}$) interaction terms cancel, but there is the second order interaction term $P_z \otimes (\sigma_2 + \sigma_{-2})$. Since it is an interaction term, it yields no contribution to the diagonal part of the Floquet operator in order $1/\omega^2$ and we can neglect it. Later on in Eq. (18) we will see that this in agreement with a higher order expansion in which this term is not neglected.

A second order effective Hamiltonian is therefore given in terms of the truncated flow equation

$$\frac{1}{\omega^2}\frac{d\mathcal{K}(l)}{dl} = -\Big(b(l) - \frac{a^2(l)}{\omega^2}\Big)P_x \otimes (\sigma_1 + \sigma_{-1}) - \frac{2}{\omega^2}a(l)b^2(l)P_z \otimes \mathbb{1}. \tag{13}$$

The solutions with boundary condition $\mathcal{K}(0) = \mathcal{K}$ satisfy

$$a(l \to \infty) = \omega_0\Big(1 - \frac{\Omega^2}{\omega^2}\Big) + O(\Omega^4/\omega^3) \text{ and} \tag{14}$$
$$b(l \to \infty) = 0 + O(\Omega^4/\omega^3) \tag{15}$$

and lead to the constant Hamiltonian

$$H_c = \omega_0\Big(1 - \frac{\Omega^2}{\omega^2}\Big)P_z + O(\Omega^4/\omega^3). \tag{16}$$

Solving the flow equations in fourth order requires adding the operators $P_z \otimes (\sigma_2 - \sigma_{-2})$, $P_y \otimes (\sigma_3 + \sigma_{-3})$ and $P_y \otimes (\sigma_1 + \sigma_{-1})$ to the set $\{\mathcal{O}_i^{(\text{int})}\}$ and taking the generator

$$\eta_4(l) = \omega b(l) P_x \otimes (\sigma_1 - \sigma_{-1}) - ia(l)b(l)P_y \otimes (\sigma_1 + \sigma_{-1}) - \frac{a(l)b^2(l)}{2\omega}P_z \otimes (\sigma_2 - \sigma_{-2})$$
$$- i\frac{a(l)b^3(l)}{6\omega^2}P_y \otimes (\sigma_3 + \sigma_{-3}) - i\frac{a(l)b^3(l)}{2\omega^2}P_y \otimes (\sigma_1 + \sigma_{-1}) \tag{17}$$

where suitable choices for the functions $c_i$ have alreay been made. With this ansatz, the flow equation reads

$$\frac{1}{\omega^2}\frac{d\mathcal{K}(l)}{dl} = -b(l)\Big(1 - \frac{a^2(l)}{\omega^2}\Big)P_x \otimes (\sigma_1 + \sigma_{-1}) - \Big(\frac{2}{\omega^2}a(l)b^2(l) + \frac{a(l)b^4(l)}{\omega^4}\Big)P_z \otimes \mathbb{1} + I(1/\omega^4), \tag{18}$$

where $I(1/\omega^4)$ are interaction terms $\sim 1/\omega^4$ that can be neglected, similarly to above, since they could be taken care of in an expansion up to $1/\omega^5$. The solutions of Eq. (18) with the proper boundary conditions satisfy

$$a(l \to \infty) = \omega_0\left(1 - \frac{\Omega^2}{\omega^2} - \frac{\omega_0^2\Omega^2}{\omega^4} + \frac{\Omega^4}{4\omega^4}\right) + O(\Omega^6/\omega^5) \,,$$
$$b(l \to \infty) = 0 + O(\Omega^6/\omega^5) \,. \tag{19}$$

Again, the interaction vanishes for $l \to \infty$, and Eq. (19) reproduces the effective level shift of a driven two-level system in fourth order [2].

## II. SHAKEN OPTICAL LATTICES

The Floquet operator of the shaken optical lattice Hamiltonian reads

$$\mathcal{K} = \left(\sum_i \mathcal{J}_0 J(c_i^\dagger c_{i+1} + c_{i+1}^\dagger c_i) + U \sum_i \hat{n}_i(\hat{n}_i - 1)\right) \otimes \mathbb{1} + \mathbb{1} \otimes \omega \hat{n}$$
$$+ \sum_{i,m \neq 0} J(\mathcal{J}_m c_i^\dagger c_{i+1} + \mathcal{J}_{-m} c_{i+1}^\dagger c_i) \otimes \sigma_m, \tag{20}$$

where $\delta^\pm(t)$, defined in Eq. (8) in the main paper, has been expressed as a Fourier series $\delta^\pm(t) = J\sum_{m=-\infty}^{\infty} \mathcal{J}_{\pm m}(K/\omega)e^{im\omega t} - J\mathcal{J}_0(K/\omega)$ with the $m$-th order Bessel function. As above, we first consider the regime of fast driving, $\{J,U\} \ll \omega$. Since experiments are typically performed in the strong driving regime with $K/\omega \gtrsim 1$, functions like $\mathcal{J}_m$ that depend on this ratio must not be expanded in a finite series, but need to be considered of order $\sim 1$.

We start our analysis with the following parametrization of the flowing Floquet operator:

$$\mathcal{K}_1(l) = \left(\sum_i \mathcal{J}_0 J(c_i^\dagger c_{i+1} + c_{i+1}^\dagger c_i) + U \sum_i \hat{n}_i(\hat{n}_i - 1)\right) \otimes \mathbb{1} + \mathbb{1} \otimes \omega \hat{n}$$
$$+ \sum_{i,m \neq 0} b_m(l) J(\mathcal{J}_m c_i^\dagger c_{i+1} + \mathcal{J}_{-m} c_{i+1}^\dagger c_i) \otimes \sigma_m. \tag{21}$$

According to our discussion above, the generator should read

$$\eta_1(l) = [\mathcal{D}, \mathcal{K}_{\text{int}}(l)] = \sum_{i,m\neq 0} \omega m b_m(l) J(\mathcal{J}_m c_i^\dagger c_{i+1} + \mathcal{J}_{-m} c_{i+1}^\dagger c_i) \otimes \sigma_m. \tag{22}$$

With this choice, however, one would obtain $U_F(t=0) \neq \mathbb{1}$. In order to avoid the additional step of introducing a constant Hamiltonian rather than identifying $H_{\text{eff}}$ directly, we choose a slightly modified generator

$$\tilde{\eta}_1(l) = J\omega \sum_{i,m\neq 0} m b_m(l)(\mathcal{J}_m c_i^\dagger c_{i+1} + \mathcal{J}_{-m} c_{i+1}^\dagger c_i) \otimes (\sigma_m - \mathbb{1}). \tag{23}$$

With this choice, the flow equation becomes

$$\frac{1}{\omega^2}\frac{d\mathcal{K}_1(l)}{dl} = 2\frac{UJ}{\omega}\sum_{i,m\neq 0} m b_m(l)\big(\mathcal{J}_m c_i^\dagger(\hat{n}_i - \hat{n}_{i+1})c_{i+1} + \mathcal{J}_{-m} c_{i+1}^\dagger(\hat{n}_{i+1} - \hat{n}_i)c_i\big) \otimes \mathbb{1}$$
$$- 2\frac{UJ}{\omega}\sum_{i,m\neq 0} m b_m(l)\big(\mathcal{J}_m c_i^\dagger(\hat{n}_i - \hat{n}_{i+1})c_{i+1} + \mathcal{J}_{-m} c_{i+1}^\dagger(\hat{n}_{i+1} - \hat{n}_i)c_i\big) \otimes \sigma_m$$
$$- J\sum_{i,m\neq 0} m^2 b_m(l)(\mathcal{J}_m c_i^\dagger c_{i+1} + \mathcal{J}_{-m} c_{i+1}^\dagger c_i) \otimes \sigma_m. \tag{24}$$

Again, the interaction terms in Eq. (24) $\sim c_{i\pm 1}^\dagger(\hat{n}_{i\pm 1} - \hat{n}_i)c_i \otimes \sigma_m$ can be neglected, as seen by using the generator

$$\tilde{\eta}_2(l) = \tilde{\eta}_1(l) - 2UJ\sum_{i,m\neq 0} b_m(l)\big(\mathcal{J}_m c_i^\dagger(\hat{n}_i - \hat{n}_{i+1})c_{i+1} + \mathcal{J}_{-m} c_{i+1}^\dagger(\hat{n}_{i+1} - \hat{n}_i)c_i\big) \otimes (\sigma_m - \mathbb{1}), \tag{25}$$

and dropping all terms of order $1/\omega^2$.

In contrast to the driven two-level system above, it is necessary to expand the set $\{\mathcal{O}_i^{(0)}\}$. Specifically, the tunneling interaction term $c_i^\dagger(\hat{n}_i - \hat{n}_{i\pm1})c_{i\pm1}$ needs to be added and will appear in the resulting effective Hamiltonian. With the extended set $\{\mathcal{O}_i^{(0)}\}$ the flowing Floquet operator reads

$$\mathcal{K}_2(l) = \mathbb{1} \otimes \omega\hat{n} + \Big(\sum_i \mathcal{J}_0 J(c_i^\dagger c_{i+1} + c_{i+1}^\dagger c_i) + U\sum_i \hat{n}_i(\hat{n}_i - 1)\Big) \otimes \mathbb{1}$$
$$+ 2\sum_i \big(a_+(l)c_i^\dagger(\hat{n}_i - \hat{n}_{i+1})c_{i+1} + a_-(l)c_{i+1}^\dagger(\hat{n}_{i+1} - \hat{n}_i)c_i\big) \otimes \mathbb{1}$$
$$+ \sum_{i,m\neq 0} b_m(l) J(\mathcal{J}_m c_i^\dagger c_{i+1} + \mathcal{J}_{-m} c_{i+1}^\dagger c_i) \otimes \sigma_m \tag{26}$$

with the boundary conditions $a_\pm(0) = 0$. Moreover, one can set $a_\pm(l) = a_\pm(0)$ on the right hand side of the flow equation $d\mathcal{K}_2/dl = [\tilde{\eta}_2, \mathcal{K}_2]$ and obtain the exact effective Hamiltonian in order $1/\omega$ after solving the flow equation

$$\frac{1}{\omega^2}\frac{d\mathcal{K}_2(l)}{dl} = -J \sum_{i,m\neq 0} m^2 b_m(l)(\mathcal{J}_m c_i^\dagger c_{i+1} + \mathcal{J}_{-m} c_{i+1}^\dagger c_i) \otimes \sigma_m$$
$$+ 2\frac{UJ}{\omega}\sum_{i,m\neq 0} m b_m(l)\big(\mathcal{J}_m c_i^\dagger(\hat{n}_i - \hat{n}_{i+1})c_{i+1} + \mathcal{J}_{-m} c_{i+1}^\dagger(\hat{n}_{i+1} - \hat{n}_i)c_i\big) \otimes \mathbb{1} + O(1/\omega^2). \tag{27}$$

The solutions of Eq. (27) yield

$$a_\pm(l \to \infty) = \pm\frac{UJ}{\omega}\beta + O(J^3/\omega^2) \text{ and} \tag{28}$$
$$b_m(l \to \infty) = 0 + O(J^3/\omega^2), \tag{29}$$

with $\beta = \sum_{m\neq 0} \frac{\mathcal{J}_m(K/\omega)}{m} = 2\sum_{m=1}^\infty \frac{\mathcal{J}_{2m-1}(K/\omega)}{2m-1}$. Consequently, $\mathcal{K}_2$ satisfies

$$\mathcal{K}_2(l \to \infty) = \Big(\sum_i \mathcal{J}_0 J(c_i^\dagger c_{i+1} + c_{i+1}^\dagger c_i) + U\sum_i \hat{n}_i(\hat{n}_i - 1) + 2\beta\frac{UJ}{\omega}\sum_i c_i^\dagger(\hat{n}_i - \hat{n}_{i+1})c_{i+1} + h.c.\Big) \otimes \mathbb{1}$$
$$+ \mathbb{1} \otimes \omega\hat{n} + O(J^3/\omega^2), \tag{30}$$

which results in the effective Hamiltonian given in Eq. (9) in the main paper. Since, as argued above, $\beta$ needs to be considered a quantity $\sim 1$, Eq. (30) is indeed of first order in $1/\omega$.

### III. LARGE INTERACTION ENERGY REGIME

Here, we show how the flow equation permits the identification of the effective Hamiltonian in first order in $J$ but in *all* orders in $U$ with little extra effort. Consider Eq. (24) and the regime $J \ll \{U, \omega\}$. Now, the generator (25) which cancels the undesired interaction in Eq. (24), creates new terms (originated from the commutator with $U\sum_i \hat{n}_i(\hat{n}_i - 1)$) of the *same order* that the interaction eliminated

$$\frac{1}{\omega^2}\frac{d\mathcal{K}_1(l)}{dl} = [\tilde{\eta}_2(l), \mathcal{K}_1(l)]$$
$$= -J\sum_{i,m\neq 0} m^2 b_m(l)(\mathcal{J}_m c_i^\dagger c_{i+1} + \mathcal{J}_{-m} c_{i+1}^\dagger c_i) \otimes \sigma_m$$
$$+ 2\frac{UJ}{\omega}\sum_{i,m\neq 0} m b_m(l)\big(\mathcal{J}_m c_i^\dagger(\hat{n}_i - \hat{n}_{i+1})c_{i+1} + \mathcal{J}_{-m} c_{i+1}^\dagger(\hat{n}_{i+1} - \hat{n}_i)c_i\big) \otimes \mathbb{1}$$
$$+ 2\frac{U^2 J}{\omega^2}\sum_{i,j,m\neq 0} b_m(l)\big(\mathcal{J}_m[c_i^\dagger(\hat{n}_i - \hat{n}_{i+1})c_{i+1}, \hat{n}_j(\hat{n}_j - 1)] + \mathcal{J}_{-m}[c_{i+1}^\dagger(\hat{n}_{i+1} - \hat{n}_i)c_i, \hat{n}_j(\hat{n}_j - 1)]\big) \otimes \mathbb{1}$$
$$- 2\frac{U^2 J}{\omega^2}\sum_{i,j,m\neq 0} b_m(l)\big(\mathcal{J}_m[c_i^\dagger(\hat{n}_i - \hat{n}_{i+1})c_{i+1}, \hat{n}_j(\hat{n}_j - 1)] + \mathcal{J}_{-m}[c_{i+1}^\dagger(\hat{n}_{i+1} - \hat{n}_i)c_i, \hat{n}_j(\hat{n}_j - 1)]\big) \otimes \sigma_m$$
$$+ O(J^2/\omega), \tag{31}$$

namely the operators $[c_i^\dagger(\hat{n}_i - \hat{n}_{i\pm 1})c_{i\pm 1}, \hat{n}_j(\hat{n}_j - 1)] \otimes (\sigma_m - \mathbb{1})$ with a coefficient $\sim JU^2/\omega^2$, which now is of order $O(J)$ and cannot be straightforwardly neglected. Modifying again the generator, one can cancel the interaction terms $[c_i^\dagger(\hat{n}_i - \hat{n}_{i\pm 1})c_{i\pm 1}, \hat{n}_j(\hat{n}_j - 1)] \otimes \sigma_m$ in the flow equation and create new terms $\sim JU^3/\omega^3$, again of order $O(J)$.

By iterating this procedure one obtains the generator

$$\eta_U(l) = \tilde{\eta}_1(l) + J\omega \sum_{n=1}^{\infty} \frac{U^n}{\omega^n} \sum_{m \neq 0} \frac{b_m(l)}{m^{n-1}} (\mathcal{J}_m \hat{C}_n^+ + \mathcal{J}_{-m} \hat{C}_n^-) \otimes (\sigma_m - \mathbb{1}), \tag{32}$$

where the operators $\hat{C}_n^\pm$ are defined recursively as follows:

$$\hat{C}_1^\pm = \Big[\sum_i c_i^\dagger c_{i\pm 1}, \sum_j \hat{n}_j(\hat{n}_j - 1)\Big] = -2\sum_i c_i^\dagger(\hat{n}_i - \hat{n}_{i\pm 1})c_{i\pm 1}, \tag{33}$$

$$\hat{C}_{n+1}^\pm = \Big[\hat{C}_n^\pm, \sum_j \hat{n}_j(\hat{n}_j - 1)\Big]. \tag{34}$$

With this generator, the flow equation becomes

$$\frac{1}{\omega^2} \frac{d\mathcal{K}_1(l)}{dl} = [\eta_U(l), \mathcal{K}_1(l)] \tag{35}$$

$$= -J \sum_{i, m \neq 0} m^2 b_m(l) (\mathcal{J}_m c_i^\dagger c_{i+1} + \mathcal{J}_{-m} c_{i+1}^\dagger c_i) \otimes \sigma_m \tag{36}$$

$$- J\frac{U}{\omega} \sum_{m \neq 0} m b_m(l) (\mathcal{J}_m \hat{C}_1^+ + \mathcal{J}_{-m} \hat{C}_1^-) \otimes \mathbb{1} \tag{37}$$

$$+ J\frac{U}{\omega} \sum_{m \neq 0} m b_m(l) (\mathcal{J}_m \hat{C}_1^+ + \mathcal{J}_{-m} \hat{C}_1^-) \otimes \sigma_m \tag{38}$$

$$- J \sum_{n=1}^{\infty} \frac{U^n}{\omega^n} \sum_{m \neq 0} \frac{b_m(l)}{m^{n-2}} (\mathcal{J}_m \hat{C}_n^+ + \mathcal{J}_{-m} \hat{C}_n^-) \otimes \sigma_m \tag{39}$$

$$+ J \sum_{n=1}^{\infty} \frac{U^{n+1}}{\omega^{n+1}} \sum_{m \neq 0} \frac{b_m(l)}{m^{n-1}} (\mathcal{J}_m \hat{C}_{n+1}^+ + \mathcal{J}_{-m} \hat{C}_{n+1}^-) \otimes \sigma_m \tag{40}$$

$$- J \sum_{n=1}^{\infty} \frac{U^{n+1}}{\omega^{n+1}} \sum_{m \neq 0} \frac{b_m(l)}{m^{n-1}} (\mathcal{J}_m \hat{C}_{n+1}^+ + \mathcal{J}_{-m} \hat{C}_{n+1}^-) \otimes \mathbb{1} \tag{41}$$

$$+ O(J^2/\omega) \tag{42}$$

Expressions (38), (39) and (40) cancel each other and the flow equation reduces to

$$\frac{1}{\omega^2} \frac{d\mathcal{K}_1(l)}{dl} = -J \sum_{i, m \neq 0} m^2 b_m(l) (\mathcal{J}_m c_i^\dagger c_{i+1} + \mathcal{J}_{-m} c_{i+1}^\dagger c_i) \otimes \sigma_m$$

$$- J \sum_{n=1}^{\infty} \frac{U^n}{\omega^n} \sum_{m \neq 0} \frac{b_m(l)}{m^{n-2}} (\mathcal{J}_m \hat{C}_n^+ + \mathcal{J}_{-m} \hat{C}_n^-) \otimes \mathbb{1} + O(J^2/\omega). \tag{43}$$

Like in the previous section, one needs to introduce a new flowing Floquet operator, $\tilde{\mathcal{K}}_2(l)$, containing the operational structure $\hat{C}_n^\pm$. Because all terms of order $O(J^2/\omega)$ will be neglected, it is sufficient to solve Eq. (57) with $\tilde{\mathcal{K}}_2(l)$ on the left-hand side of the equation instead of $\mathcal{K}_1(l)$. The result of the differential equation yields the effective Hamiltonian

$$H_{\text{eff}}^U = \sum_i \mathcal{J}_0 J(c_i^\dagger c_{i+1} + c_{i+1}^\dagger c_i) + U \sum_i \hat{n}_i(\hat{n}_i - 1) - \sum_{n=1}^{\infty} J\frac{U^n}{\omega^n} (\beta_n^+ \hat{C}_n^+ + \beta_n^- \hat{C}_n^-) + O(J^2/\omega), \tag{44}$$

with

$$\beta_n^\pm = \sum_{m \neq 0} \frac{\mathcal{J}_{\pm m}(K/\omega)}{m^n} = \sum_{m \neq 0} (\pm 1)^n \frac{\mathcal{J}_m(K/\omega)}{m^n}. \tag{45}$$

## IV. LARGE TUNNELING AMPLITUDE REGIME

Consider now the large hopping matrix element regime, $U \ll \{J, \omega\}$. Similarly to above, one obtains the generator

$$\eta_t(l) = \tilde{\eta}_1(l) + U\omega \sum_{n=1}^{\infty} \frac{J^n}{\omega^n} \sum_{m \neq 0} \frac{b_m(l)}{m^{n-1}} \mathcal{J}_0^{n-1}(\mathcal{J}_m \hat{T}_n^+ + \mathcal{J}_{-m} \hat{T}_n^-) \otimes (\sigma_m - \mathbb{1}), \tag{46}$$

where the operators $\hat{T}_n^{\pm}$ are defined recursively as follows:

$$\hat{T}_1^{\pm} = \Big[\sum_i c_i^{\dagger} c_{i\pm 1}, \sum_j \hat{n}_j(\hat{n}_j - 1)\Big] = -2 \sum_i c_i^{\dagger}(\hat{n}_i - \hat{n}_{i\pm 1}) c_{i\pm 1}, \tag{47}$$

$$\hat{T}_{n+1}^{\pm} = \Big[\hat{T}_n^{\pm}, \sum_j (c_j^{\dagger} c_{j+1} + c_{j+1}^{\dagger} c_j)\Big]. \tag{48}$$

The flow equation becomes

$$\frac{1}{\omega^2} \frac{d\mathcal{K}_1(l)}{dl} = [\eta_t(l), \mathcal{K}_1(l)] \tag{49}$$

$$= -J \sum_{i, m \neq 0} m^2 b_m(l)(\mathcal{J}_m c_i^{\dagger} c_{i+1} + \mathcal{J}_{-m} c_{i+1}^{\dagger} c_i) \otimes \sigma_m \tag{50}$$

$$- U \frac{J}{\omega} \sum_{m \neq 0} m b_m(l)(\mathcal{J}_m \hat{T}_1^+ + \mathcal{J}_{-m} \hat{T}_1^-) \otimes \mathbb{1} \tag{51}$$

$$+ U \frac{J}{\omega} \sum_{m \neq 0} m b_m(l)(\mathcal{J}_m \hat{T}_1^+ + \mathcal{J}_{-m} \hat{T}_1^-) \otimes \sigma_m \tag{52}$$

$$- U \sum_{n=1}^{\infty} \frac{J^n}{\omega^n} \sum_{m \neq 0} \frac{b_m(l)}{m^{n-2}} \mathcal{J}_0^{n-1}(\mathcal{J}_m \hat{T}_n^+ + \mathcal{J}_{-m} \hat{T}_n^-) \otimes \sigma_m \tag{53}$$

$$+ U \sum_{n=1}^{\infty} \frac{J^{n+1}}{\omega^{n+1}} \sum_{m \neq 0} \frac{b_m(l)}{m^{n-1}} \mathcal{J}_0^n (\mathcal{J}_m \hat{T}_{n+1}^+ + \mathcal{J}_{-m} \hat{C}_{n+1}^-) \otimes \sigma_m \tag{54}$$

$$- U \sum_{n=1}^{\infty} \frac{J^{n+1}}{\omega^{n+1}} \sum_{m \neq 0} \frac{b_m(l)}{m^{n-1}} \mathcal{J}_0^n (\mathcal{J}_m \hat{T}_{n+1}^+ + \mathcal{J}_{-m} \hat{T}_{n+1}^-) \otimes \mathbb{1} \tag{55}$$

$$+ O(U^2/\omega). \tag{56}$$

Expressions (52), (53) and (54) cancel each other and the flow equation reduces to

$$\frac{1}{\omega^2} \frac{d\mathcal{K}_1(l)}{dl} = -J \sum_{i, m \neq 0} m^2 b_m(l)(\mathcal{J}_m c_i^{\dagger} c_{i+1} + \mathcal{J}_{-m} c_{i+1}^{\dagger} c_i) \otimes \sigma_m$$
$$- U \sum_{n=1}^{\infty} \mathcal{J}_0^{n-1} \frac{J^n}{\omega^n} \sum_{m \neq 0} \frac{b_m(l)}{m^{n-2}} (\mathcal{J}_m \hat{T}_n^+ + \mathcal{J}_{-m} \hat{T}_n^-) \otimes \mathbb{1} + O(U^2/\omega). \tag{57}$$

Finally, the effective Hamiltonian becomes

$$H_{\text{eff}}^J = \sum_i \mathcal{J}_0 J(c_i^{\dagger} c_{i+1} + c_{i+1}^{\dagger} c_i) + U \sum_i \hat{n}_i(\hat{n}_i - 1) - U \sum_{n=1}^{\infty} \mathcal{J}_0^{n-1} \frac{J^n}{\omega^n} (\beta_n^+ \hat{T}_n^+ + \beta_n^- \hat{T}_n^-) + O(U^2/\omega), \tag{58}$$

where $\beta_n^{\pm}$ are defined in Eq. (45).

---


\end{widetext}

\end{document}